\documentclass{iopart}

\usepackage{graphicx}
\usepackage{dcolumn}
\usepackage{textcomp}

\begin{document}

\title{Spatial conductivity mapping of unprotected and capped black phosphorus using microwave microscopy} 

\author{Pieter J. de Visser$^{1,2}$, Rebekah Chua$^{1,3}$, Joshua O. Island$^{1}$, Matvey Finkel$^{1,4}$, Allard J. Katan$^{1}$, Holger Thierschmann$^{1}$, Herre S. J. van der Zant$^{1}$ and Teun M. Klapwijk$^{1,4}$}
\address{$^1$Kavli Institute of NanoScience, Faculty of Applied Sciences, Delft University of Technology, Lorentzweg 1, 2628 CJ Delft, The Netherlands}
\address{$^2$Department of Quantum Matter Physics, University of Geneva, Geneva 1211, Switzerland}
\address{$^3$Department of Physics, National University of Singapore, 2 Science Drive 3, Singapore 117542, Singapore}
\address{$^4$Physics Department, Moscow State Pedagogical University, Moscow 119991, Russia}

\date{\today}

\begin{abstract}

Thin layers of black phosphorus present an ideal combination of a 2D material with a tunable direct bandgap and high carrier mobility. However the material suffers from degradation in ambient conditions due to an oxidation reaction which involves water, oxygen and light. We have measured the spatial profile of the conductivity on flakes of black phosphorus as a function of time using scanning microwave impedance microscopy. A microwave excitation (3 GHz) allows to image a conducting sample even when covered with a dielectric layer. We observe that on bare black phosphorus, the conductivity changes drastically over the whole surface within a day. We demonstrate that the degradation process is slowed down considerably by covering the material with a 10 nm layer of hafnium oxide. It is stable for more than a week, opening up a route towards stable black phosphorus devices in which the high dielectric constant of hafnium oxide can be exploited. Covering black phosphorus with a 15 nm boron nitride flake changes the degradation process qualitatively, it is dominated by the edges of the flake indicating a diffusive process and happens on the scale of days. 

\end{abstract}

\maketitle

Black phosphorus is a solid, Van der Waals bonded form of elemental phosphorus which can be synthesised under high pressure. A hundred years after the discovery of the bulk form and subsequent investigation of the bulk properties, black phosphorus (black P) was very recently rediscovered as a two-dimensional layered material \cite{lli2014}. Unlike the other elemental 2D material graphene, black P has a direct bandgap, which is tuneable with layer thickness from around 2 eV for a monolayer to 0.3 eV for bulk \cite{vtran2014,sdas2014}. Combined with a high mobility, it is a very promising material for optoelectronic applications. Several reviews appeared this year that summarise and provide references to the rapidly growing body of literature on the chemistry, physics and applications of black P \cite{xling2015,hdu2015,hliu2015,acastellanosgomez2015}. 

For any application in ambient conditions the environmental stability of a material is crucial. However black P has been shown to degrade very rapidly without protective measures. Various studies have been carried out on water absorption at the hydrophilic surface of black P and the consequences for devices \cite{skoenig2014,acastellanosgomez2014b,jisland2015}. Chemical analysis shows that degradation occurs due to photooxidation, a process which simultaneously requires oxygen, water and energy (light) \cite{afavron2015,aziletti2015} and ends with a transformation to phosphoric acid. The reaction rate depends exponentially on the square of the energy gap and therefore depends strongly on the number of layers, and could be initiated on material defects \cite{kutt2015}. 

Attempts to protect black P from environmental degradation in ambient conditions involve capping layers of deposited AlO$_\mathrm{x}$ \cite{jwood2014,xluo2014,jna2014,wzhu2015,nyoungblood2015}, other 2D materials such as boron nitride (BN) \cite{aavsar2015,rdoganov2015,xchen2015,ngillgren2015,lli2015,ycao2015}, polymer layers \cite{vtayari2015} or combinations of these \cite{acastellanosgomez2015,jkim2015}. Where there are many techniques to monitor the degradation process for a bare black P flake (e.g. optical microscopy, atomic force microscopy (AFM), Raman scattering, chemical analysis), the techniques to monitor the effectiveness of a capping layer are scarce. Most of the cited studies monitor the degradation under a capping layer by the performance (conductivity) of a transistor channel, which reduces the information to a single number as a function of time. This is useful to characterise device stability, but does hide the spatial conductivity profile of the flake while (possibly) degrading. 

Here we report scanning Microwave Impedance Microscopy (sMIM) measurements on bare black P, black P protected with ALD-deposited HfO$_2$ and black P protected with BN. sMIM is an AFM-based microscope technique, which measures the local conductivity of a sample by monitoring the microwave reflection from a metallic tip with a typical resolution of $\sim$50 nm \cite{jkim2015,klai2009,klai2010}. Because the high frequency signal (3 GHz) penetrates through dielectric layers, the conductivity of a buried layer below a dielectric can be imaged without the need for contacts. We show that on a bare black P flake, the conductivity changes drastically over the whole surface over the course of hours, consistent with the previously discussed literature. We demonstrate that the degradation process is slowed down considerably by covering the material with a 10 nm layer of hafnium oxide, under which black P remains stable for at least 11 days. This opens up a route towards stable black P devices with HfO$_2$, in which the high dielectric constant of HfO$_2$ can be used (e.g. as a gate dielectric). A recent studies also suggests that the mobility in BP is enhanced due to a HfO$_2$ capping layer \cite{venkatakamalakar2015}. Covering black phosphorus with a 15 nm boron nitride flake does not completely stabilise black P and changes the phenomenology of the degradation process compared to the bare flake. It is much slower (days) and is dominated by the edges of the black P flake. This indicates that the surface is protected, but a diffusive process still degrades the black P starting from the edges despite the full coverage with a large BN flake. Although quite a few studies on capping black P with boron nitride have been reported, we present the first conductivity images of the buried black P under BN as a function of time. Signatures of such an edge process were also found in black P covered with thin AlO$_\mathrm{x}$ \cite{jkim2015} and in solution processed black P \cite{dhanlon2015}.  

Preparation of few-layer black P flakes is carried out using a modified mechanical exfoliation technique as discussed in detail in \cite{acastellanosgomez2014}. Bulk, commercially available black P (99.998\%, Smart Elements) is used in this study. By employing a viscoelastic stamp (poly- dimethilsiloxane (PDMS)) as an intermediate substrate for exfoliation, thin black P flakes are transferred to a SiO$_2$ substrate leaving less residue compared with using Scotch tape only. Thin flakes are then identified by contrast under an optical microscope. We have measured samples using three different ways of preparation as described below. We monitored each sample with sMIM over time. The first sample contains bare black P flakes. For the second sample, after preparing the black P flakes on SiO$_2$, the whole sample is covered with a layer of 10 nm of HfO$_\mathrm{2}$ using atomic layer deposition. For the third sample, immediately after identifying a thin flake of black P, this flake was covered with a large, thin flake (about 25 $\mu$m x 25 $\mu$m x 15 nm) of boron nitride, which completely covers the black P flake (4x6 $\mu$m). The boron-nitride flake is prepared in the same way as the black P flake onto the viscoelastic stamp and positioned on top of the selected black P flake under an optical microscope to make sure the latter is fully covered. Detailed characterisation using optical microscopy, atomic force microscopy, transport measurements, Raman spectroscopy and transmission electron microscopy of the same black P using the same exfoliation technique was reported before in References \cite{acastellanosgomez2014b} and \cite{jisland2015}.

\begin{figure}
\includegraphics[width=0.99\columnwidth]{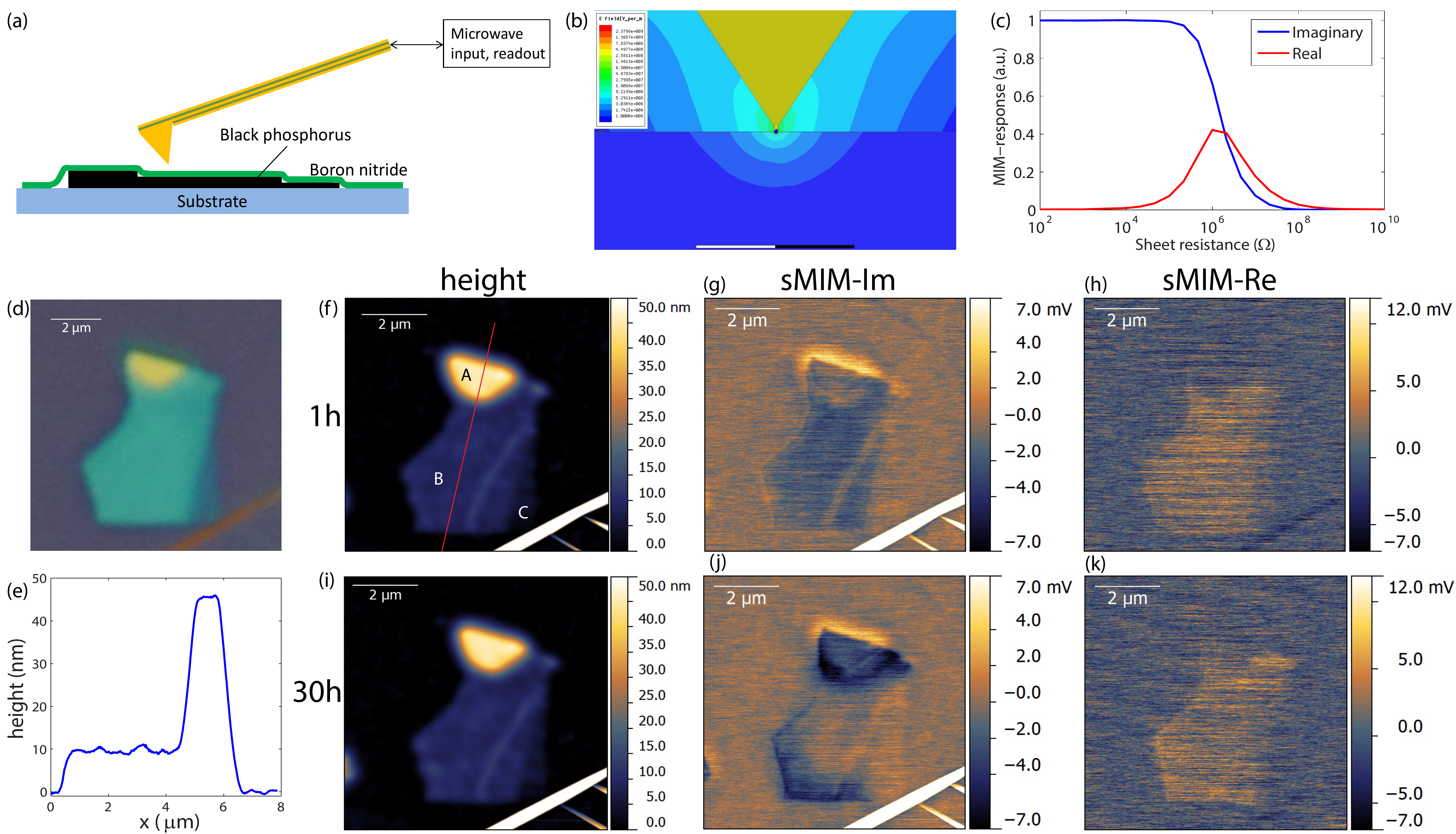}
\caption{\label{fig:sMIM_Fig1} (a) Schematic of a black P flake on top of a SiO$_2$ substrate, covered by a larger flake of BN. The sMIM cantilever is sensitive to local conductivity variations. (b) Simulated electric field distribution in the vicinity of the tip apex. The scale bar at the bottom represents a length of 5 $\mu$m. (c) Simulated response curve of the sMIM system as a function of sheet resistance of the flake under study. (d) Optical microscope image of the BN covered black P flake. (e) AFM height profile of the BN covered black P flake, at the position indicated with the red line in (f). (f) AFM height map of the flake showing three regions with different thickness (46 nm, 9 nm and 5 nm), indicated as A, B and C. (g,h) The imaginary and real part of the microwave response respectively 1 hour after the fabrication of the sample. (i-k) Height profile, imaginary and real part of the microwave response after 30 hours.}
\end{figure}

To measure the time dependence of the local sheet resistance of the black P flakes we employed sMIM. This technique is based on a microwave signal (3 GHz) which is launched to a shielded, nanofabricated cantilever (Primenano) using a coaxial cable. The reflected signal contains information about the local tip-sample impedance. The typical spatial resolution is given by the tip radius, nominally 50 nm. The microwave system (ScanWave, Primenano) is mounted on a regular AFM (Cypher, Asylum Research). sMIM therefore provides topographical information of the sample, combined with a map of local conductivity variations. It is ideal to study the local temporal conductivity variations of black P since it does not require contacts, it allows for flakes with thickness variations, and it can image through dielectric layers such as the BN (in contrast to DC-transport based techniques). A sketch of the sMIM technique is shown in Figure \ref{fig:sMIM_Fig1}a.

\begin{figure}
\includegraphics[width=0.8\columnwidth]{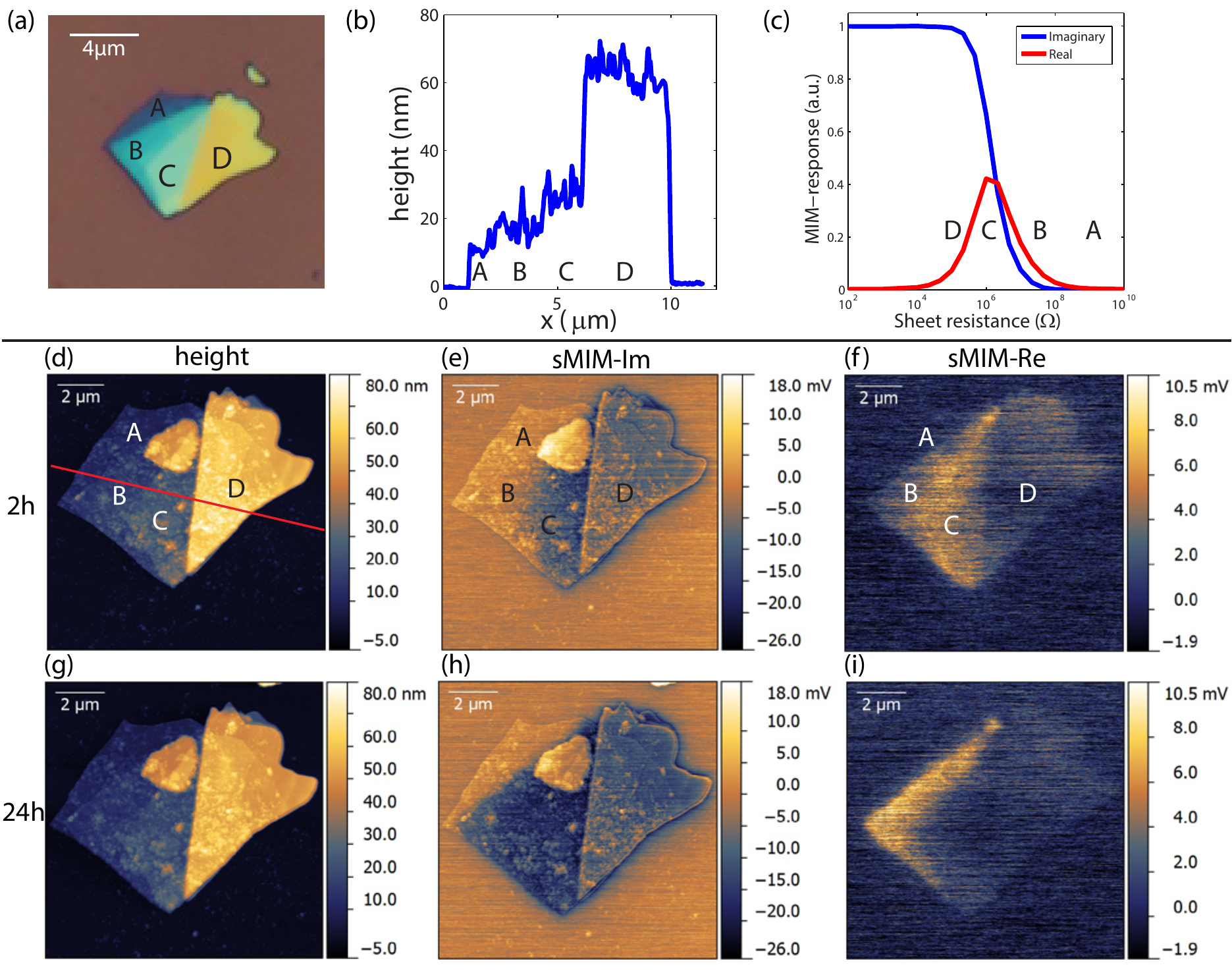}
\caption{\label{fig:time_BPbare} (a) Optical microscope image of a bare black phosphorus flake. The letters A-D indicate the regions with different thickness (8 nm, 13 nm, 20 nm, and 55 nm respectively). (b) Height profile of the flake along the line cut indicated in red in panel (d). (c) Simulated response curve of the sMIM system as a function of sheet resistance of the flake under study. The letters A-D indicate the approximate sheet resistance of the regions with different thickness marked in the images. (d-f) Height map, and the imaginary and real part of the microwave response after 2 hours. (g-i) Height map, and the imaginary and real part of the microwave response after 24 hours.}
\end{figure}

\begin{figure}
\includegraphics[width=0.99\columnwidth]{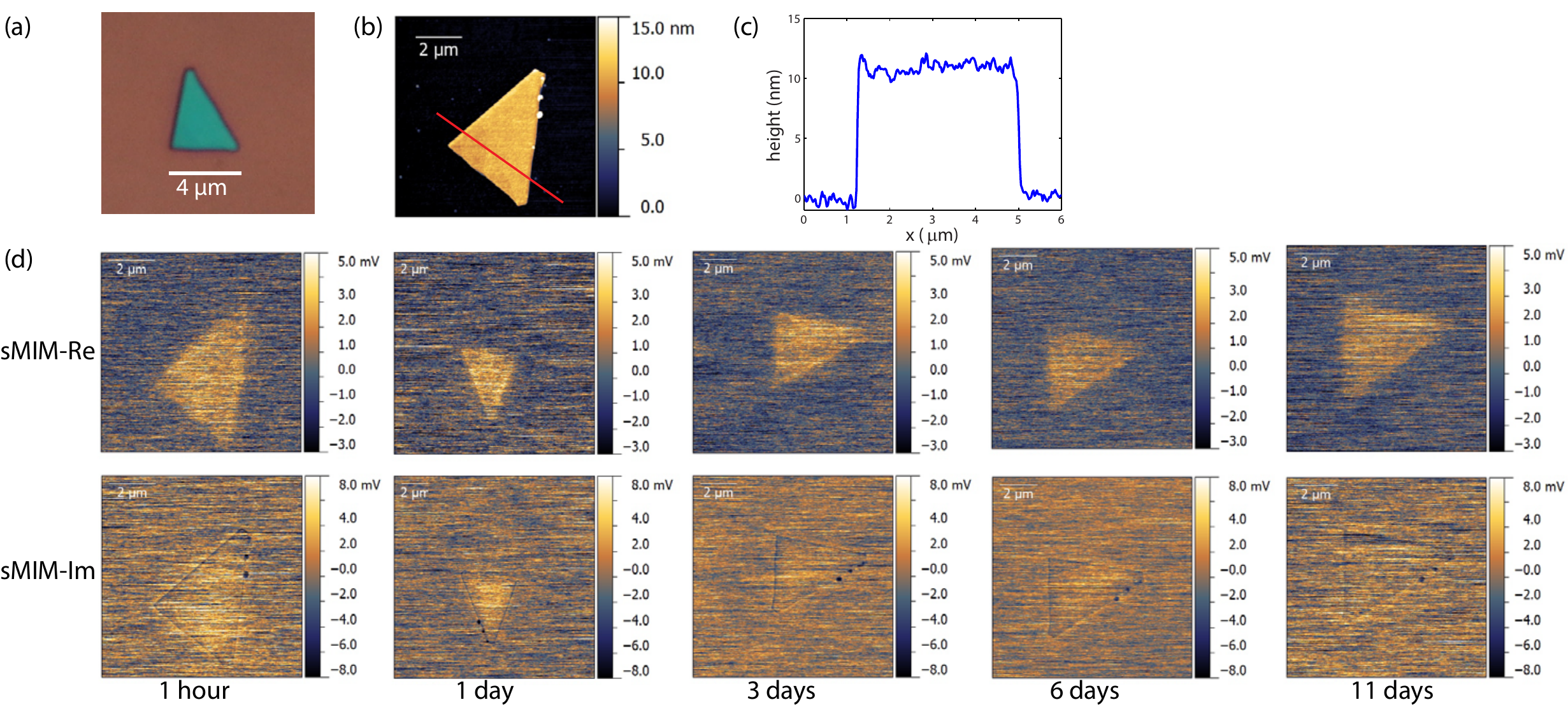}
\caption{\label{fig:time_HfOx} (a) Optical microscope image of a black P flake, which is covered by 10 nm layer of HfO$_2$. (b) Height map. (c) Height profile along the line shown in red in panel (b). (d) Real (top row) and imaginary part (bottom row) of the microwave response as a function of time. The time after fabrication of the sample is shown below the images.}
\end{figure}

\begin{figure}
\includegraphics[width=0.99\columnwidth]{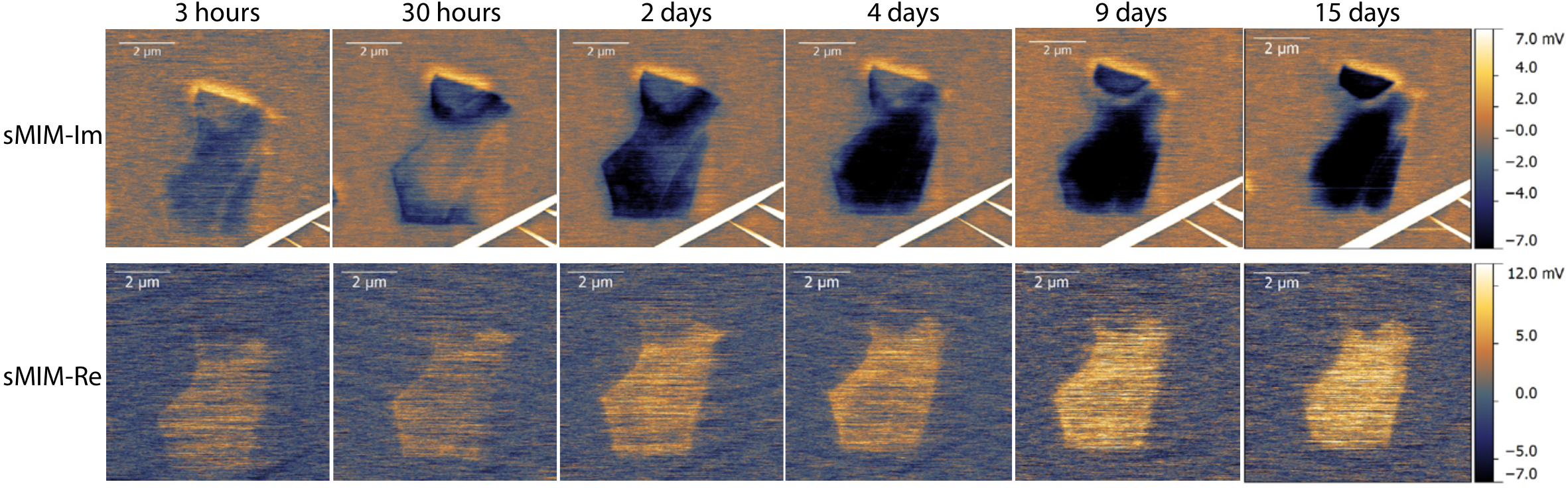}
\caption{\label{fig:time_BPBN} Imaginary (top row) and real part (bottom row) of the microwave response of the BN-covered black P flake as a function of time. The same colour scale applies to all panels. The time after fabrication of the sample is shown above the pictures. }
\end{figure}

To relate the microwave response to the local sheet resistance of the sample, we have performed finite element simulations using HFSS. Above the tip apex, the metallic part extends in pyramidal form to 5 $\mu$m above the sample, where it ends in the central strip of the (shielded) microwave stripline on the cantilever. Therefore in the simulation the whole tip (pyramid + apex) has to be taken into account. We define a port at the point where the transmission line starts and calculate the complex reflection coefficient as a function of the sheet resistance of the sample underneath the tip. The tip-sample distance is taken to be 0.2 nm. The simulated sample is a 5x5 $\mu$m disk on an insulating substrate with a dielectric constant of 3.9 (SiO$_2$). The resulting electric field distribution is shown in the colour plot in Figure \ref{fig:sMIM_Fig1}b. The real and imaginary part of the microwave response are shown in Figure \ref{fig:sMIM_Fig1}c. The real part corresponds to resistance (losses) and the imaginary part to the reactance (capacitance), which both change if the sheet resistance is varied. We have simulated all three cases (bare, with BN, with HfO$_2$) in HFSS and found the same response curves. We use the maximum contrast between the real and imaginary response at low sheet resistance to frequently calibrate the response, using a sample with metal structures on an insulating substrate. We then calibrate the measured data to account for long term changes of the tip (due to e.g. wear) in between the measurements in which we are monitoring the time dependence of the sheet resistance of the sample. The maximum signal on the calibration sample is -15 mV for the data in Figures 
\ref{fig:sMIM_Fig1} and \ref{fig:time_BPBN}, -20 mV for the data in Figure \ref{fig:time_BPbare} and +8 mV for the data in Figure \ref{fig:time_HfOx}. The imaginary part of the response can be either positive or negative on the calibration sample. The sign depends on a phase-shift of the signal which varies from tip to tip, which can easily be determined from the calibration measurement. 

To demonstrate the power of the technique, we first present results on the BN capped sample. In this experiment we operated the AFM/sMIM in contact mode. This does not affect the degradation process because the BP flake is protected by the BN layer and at the same time gives slightly better signal to noise for the microwave signal. An optical microscope picture and an AFM height map are shown in Figures \ref{fig:sMIM_Fig1}d,f. We identify three regions with different thickness, 46 nm, 9 nm and 5 nm. A height profile crossing the first two regions is shown in Figure \ref{fig:sMIM_Fig1}e. Note that the height steps are less sharp than for a bare black P sample (in Figure \ref{fig:time_BPbare}) due to the BN flake on top. Figures \ref{fig:sMIM_Fig1}f,i show the height maps 1 hour and 30 hours after fabrication respectively, showing no noticeable change. Panels g and j compare the imaginary part of the microwave response and panels h and k the real part. Especially in the imaginary part, a drastic change is seen over time, which is the largest at the edges of the covered black P flake. The change is consistent with an increase in the conductivity as reported from transport studies. sMIM provides the additional information of how the degradation process evolves over space and time as it allows to image through the dielectric cap with high spatial resolution. Before we present the detailed time series of the BN-capped sample, we first discuss results on a bare black P flake and on a black P flake covered with HfO$_2$.

Figure \ref{fig:time_BPbare}a shows an optical microscope image of a bare black P flake, from which we identify four regions with different thickness. In this experiment we use the AFM/sMIM in tapping mode, not to disturb the formation of water droplets on the BP \cite{jisland2015}. Figure \ref{fig:time_BPbare}d shows the height map of the flake, a cross section of which is shown in Figure \ref{fig:time_BPbare}b. The indicated regions A, B, C and D have thicknesses of 8 nm, 13 nm, 20 nm, and 55 nm respectively. Figure \ref{fig:time_BPbare}c shows the simulated microwave sMIM response as a function of the sheet resistance of the studied flake. Figures \ref{fig:time_BPbare}e,f show the imaginary and real part of the microwave response right after the fabrication of the sample (within 2 hours). We observe a clear difference in response for the different thicknesses. Using both response pictures we can get an order of magnitude estimate of the sheet resistance of the regions, which is indicated in the response curve in Fig. \ref{fig:time_BPbare}c. The thinner regions have a higher sheet resistance, as expected. Thinner BP has a higher bandgap \cite{acastellanosgomez2014b}, which due to the lower number of thermally excited charge carriers leads to a lower conductivity and higher sheet resistance. On top of that a thinner layer gives a higher sheet resistance for the same conductivity. Note that the real part of the response is non-monotonic and would make it difficult to derive sheet resistance, $R_{s}$, from the microwave response, especially for regions B and C. Also the height difference by itself changes the contrast in the imaginary signal slightly (small change in capitance). Since the real part sometimes shows the largest change over time, we show here the microwave signal, not converted to sheet resistance, noting that this could in principle be done using Figure \ref{fig:time_BPbare}c.

Figures \ref{fig:time_BPbare}g-i show the height profile and microwave signal after 24 hours. We observe that for regions B and C the sheet resistance drops drastically over the course of one day, consistent with earlier transport measurements \cite{jisland2015} on similar flakes. The response of regions A and D hardly changes because their $R_s$ is in a region where the microwave response is not sensitive (Figure \ref{fig:time_BPbare}c). We also observe that the sheet resistance changes uniformly over each region with the same height, indicative of a surface effect, where the whole surface is exposed to the photo-oxidation reaction of the black P.

To study the effectiveness of an ALD-deposited dielectric, we measured the conductivity map of a sample with black P flakes, covered with a 10 nm layer of HfO$_2$ fabricated as described above. In this experiment we used the tapping mode of the AFM, because it appeared that in contact mode the tip easily picks up particles from the deposited dielectric. Figure \ref{fig:time_HfOx}a shows an optical microscope picture of a uniform, 12 nm thick, black P flake covered by HfO$_2$. Figure \ref{fig:time_HfOx}b shows the height map of the flake, with a cross section at the position indicated in red in panel c. Figure \ref{fig:time_HfOx}d shows the real (top row) and the imaginary part (bottom row) of the microwave response, as a function of time after fabrication as indicated below the images. We observe that the drastic changes in the sMIM response of the bare black P, starting on the whole surface (Figure \ref{fig:time_BPbare}), and of the BN covered black P, starting from the edges (Figure \ref{fig:time_BPBN}), do not occur. The signal on this sample is stable over the course of 11 days. This indicates that the chemical and electrical properties of black P are preserved under HfO$_2$. Compared to the Al$_2$O$_3$ cover in \cite{jkim2015} (a few days stability), this deposited dielectric preserves black P much longer. After 11 days, the signal became fainter for this flake, which we could not directly relate to either the sample or the tip, but it does not show the same pronounced degradation signatures as in the bare black P or the BN-covered case. We obtain similar results on two other flakes on the same substrate. On a higher flake (30 nm), we observed bubbles on the flake after 42 days, indicating that the HfO$_2$ slows down the degradation process, but does not completely preserve black P in the long run.

Where the results on bare black P and a deposited HfO$_2$ cover are consistent with previous works, the spatial profile of the conductivity under a BN flake capping as a function of time has not yet been reported. In presenting the sMIM technique around Figure \ref{fig:sMIM_Fig1}, we already introduced the measurement with the BN cover. Figure \ref{fig:time_BPBN} shows the time evolution of the microwave signal in the imaginary (top row) and real (bottom row) parts. We observe that initially we have low contrast (signature of a relatively high sheet resistance). However, after one day the sheet resistance drops as indicate by the stronger signal in the imaginary response, where the effect is the largest on the edges. The effect slowly diffuses inwards and only after a week we start to see the same effect on the thicker part of the sample. This process is markedly different from the bare black P data in Figure \ref{fig:time_BPbare}, where a clear surface effect is observed. Clearly the BN does not shield the black P sufficiently well from the environment. It is unlikely that water captured during fabrication causes this process, since that would rather give a homogeneous effect over the whole surface. Moreover from the bare black P results, captured water during fabrication is expected to react much quicker with black P than we observe in this experiment. A degradation process dominated by the edges was also observed in \cite{jkim2015} for a (too) thin layer of ALD deposited Al$_2$O$_3$. In \cite{jkim2015} it was argued that the sidewalls of a relatively high flake are not covered by a thin ALD deposited layer, effectively creating gaps in the protecting layer at the flakes edges. However the large BN flake that we use is a continuous cover of the black P flake and its edges, which makes such edge leakage unlikely.

We emphasise that the height of the BN covered black P does not change over the measurement time, in contrast to bare BP flakes, where a layer of water is visible on top of the flake \cite{jisland2015}. By monitoring the electronic properties of the material with sMIM, the ageing process can still be identified.

In summary, we have measured the spatial profile of the conductivity on flakes of black P with varying thickness as a function of time using sMIM. On a bare flake, the conductivity changes drastically over the whole surface over the course of hours due to the reaction of black P with water and oxygen. We demonstrate that the degradation process is slowed down to a timescale of weeks when the black P is covered with a 10 nm layer of ALD deposited HfO$_2$. This does not only extend the variety of materials with which BP can be protected, but also opens up a route towards stable black P devices in which the high dielectric constant of HfO$_2$ can be exploited. The deposition of HfO$_2$ could also be preceded by surface oxidation step of the black P, which creates a native oxide layer which potentially also protects against environmental degradation \cite{gwang2015,medmonds2015}. 

Covering black phosphorus with a 15 nm boron nitride flake changes the degradation process qualitatively, it is much slower than on a bare flake (days) and is dominated by the edges of the flake indicating a diffusive process, despite the full coverage of the black P with the BN sheet. This measurement demonstrates the unique capability of sMIM by spatially mapping the conductivity of a buried layer. 

We would like to thank Keji Lai and Eric Ma for useful discussions. T.M. Klapwijk acknowledges financial support from the Ministry of Science and Education of Russia under Contract No. 14.B25.31.0007 and from the European Research Council Advanced Grant No. 339306 (METIQUM). P.J. de Visser acknowledges support from a Niels Stensen Fellowship. J.O. Island and H.S.J. van der Zant acknowledge support by the Dutch Organization for Fundamental Research on Matter (NWO/OCW).

\section*{References}
\bibliographystyle{iopart-num}

\end{document}